\documentclass[12pt,preprint]{aastex}
\usepackage{amsmath,amsthm,amssymb}
\shorttitle{Light element production in SNe Ic}
\shortauthors{Nakamura et al.}

\def\Msun{~M_{\odot} }

\begin{document}

\title{Light Element Production in the Circumstellar Matter of Energetic Type Ic Supernovae}

\author{Ko Nakamura$^{1,2}$}
\author{Susumu Inoue$^3$}
\author{Shinya Wanajo$^1$}
\and
\author{Toshikazu Shigeyama$^1$}
\affil{$^1$Research Center for the Early Universe, Graduate School of Science, 
        $^2$Department of Astronomy, Graduate School of Science, 
        University of Tokyo, Bunkyo-ku, Tokyo 113-0033, Japan\\
        $^3$National Astronomical Observatory of Japan, Mitaka, Tokyo 181-8588, Japan
       }

\begin{abstract}
We investigate energetic type Ic supernovae as production sites for $^6$Li and Be
in the early stages of the Milky Way.
Recent observations have revealed that some very metal-poor stars with [Fe/H]$<-2.5$ possess unexpectedly high abundances of $^6$Li. Some also exbihit enhanced abundances of Be as well as N.
From a theoretical point of view, recent studies of the evolution of metal-poor massive stars
show that rotation-induced mixing can enrich the outer H and He layers with C, N, and O (CNO) elements, particularly N, and at the same time cause intense mass loss of these layers.
Here we consider energetic supernova explosions occurring after the progeniter star has lost
all but a small fraction of the He layer.
The fastest portion of the supernova ejecta can interact directly with the circumstellar matter (CSM),
both composed of He and CNO, and induce light element production through spallation and He-He fusion reactions.
The CSM should be sufficiently thick to energetic particles so that the interactions terminate
within its innermost regions.
We calculate the resulting $^6$Li/O and $^9$Be/O ratios in the ejecta$+$CSM material
out of which the very metal-poor stars may form.
We find that they are consistent with the observed values
if the mass of the He layer remaining on the pre-explosion core is $\sim 0.01-0.1 \Msun$,
and the mass fraction of N mixed in the He layer is $\sim 0.01$.
Further observations of $^6$Li, Be and N at low metallicity should provide critical tests
of this production scenario.
\end{abstract}


\section{Introduction}
Among the light elements Li, Be and B (LiBeB),
$^7$Li is thought to arise from a variety of processes, including
big bang nucleosynthesis \citep{Spite82}, asymptotic giant branch stars, novae \citep{DAntona91},
and the $\nu$-process in type II supernovae \citep{Woosley90}; the latter may also contribute to $^{11}$B.
On the other hand, the main production channel for the rest,
in particular for $^6$Li and Be, is believed to be cosmic-ray induced nuclear reactions.
The most widely discussed models of LiBeB production are based on cosmic rays accelerated in supernova shocks
\citep{Meneguzzi71,Vangioni00}.
Observations of metal-poor stars in the Galactic halo
show a primary relation between [Fe/H] and [Be/H] or [B/H],
which is consistent with spallation by cosmic rays enriched with C, N, and O (CNO) from fresh SN ejecta
impinging on interstellar H or He \citep{Duncan92}.

An alternative possibility was proposed by \citet{Fields96, Fields02},
who considered explosions of Type Ic supernovae (SNe Ic) as a site for primary LiBeB production.
Since it is expected that a fraction of C and O in the surface layers of the ejecta
are accelerated to energies above the threshold for spallation reactions when the shock passes through the stellar surface,
LiBeB production can occur through the direct interaction of the ejecta with the ambient material,
without the need for shock acceleration of cosmic rays.
This was explored in greater depth by \citet{Nakamura04},
who used more realistic stellar models and equations of state,
together with a 1-dimensional relativistic hydrodynamic code to investigate in detail
how much of the ejecta mass acquires sufficient energies for the LiBeB production.
All of these studies used stellar models which completely lost their H and He envelopes
and assumed the target to be interstellar matter (ISM) consisting of 90\% H and 10\% He,
ignoring any circumstellar matter (CSM).
Therefore, the only reactions under consideration were ${\rm C, O} + {\rm H, He} \rightarrow {\rm LiBeB}$.

Recent observations by VLT/UVES \citep{Asplund05} and Subaru/HDS (Inoue  et al. 2005, Aoki et al, in prep.)
have revealed that some very metal-poor stars possess surprisingly high abundances of $^6$Li,
much higher than expected from standard supernova cosmic ray models \citep{Ramaty00,Suzuki01,Prantzos06}.
The measured values are also higher than can be accommodated in the SN Ic production scenario discussed above,
even in the case of an energetic explosion similar to SN 1998bw, as can be seen from the results of \citet{Nakamura04}.
However, besides spallation of C and O, the fusion reaction ${\rm He + He} \rightarrow {\rm ^6Li}$ 
may also be potentially important.
This reaction will be significant in the conceivable case that
a small fraction of He is still remaining in the surface layer of the core at the time of the explosion,
while most of the He has been transported to the CSM through mass loss.
\citet{Meynet02} and \citet{Meynet06} have recently calculated the evolution of metal-poor massive stars
taking into account the effects of rotation-induced mixing, and shown that extensive mass loss occurs
even in extremely metal-poor cases,
in addition to significant enrichment of the N abundance near the surface.
If this N is accelerated at the shock breakout of the SN explosion,
a significant amount of Be can be produced through the reaction N + He $\rightarrow$ $^9$Be,
because of its low threshold $E_{\rm th}$ ($\sim 6$ MeV/A) and high cross section at peak $\sigma_{\rm p}$ ($\sim 24$ mb)
compared with other $^9$Be-producing reactions
(eg. $E_{\rm th} \sim 8$ MeV/A and $\sigma_{\rm p} \sim 8$ mb for O + He $\rightarrow$ $^9$Be, see Fig. \ref{fig-Ecs}).
Indeed, recent observations indicate that the abundances of both Be and N may be enhanced
in some metal-poor stars in which $^6$Li is detected \citep{Primas00a, Primas00b,Israelian04}.
In this Letter,
we focus on the interactions ${\rm He,CNO} + {\rm He, N} \rightarrow {\rm LiBeB}$
occurring between the ejecta accelerated by the energetic SN explosion and the CSM,
in order to account for the abundances in such very metal-poor stars.

We summarize recent observations of the relevant elements in very metal-poor stars in \S \ref{sec-obs}.
In \S \ref{sec-SNIC} we discuss SN explosions of stars with stripped He layers embedded in the CSM.
Our calculations and results are shown in \S 4 with an appropriate parameter range.
Conclusions are presented in \S \ref{sec-disc}.

\section{Summary of Observations} \label{sec-obs}
Observational determination of the abundances of $^6$Li in metal-poor stars is extremely challenging
due to its deficiency and the proximity of its absorption lines to the much stronger ones of the $^7$Li isotope.
Owing to the advent of large telescopes and the improved capabilities of observational instruments,
reliable measurements have become possible only in the last several years \citep[for earlier observations, see][]{Smith98,Hobbs99,Cayrel99}. 
\citet{Asplund05} observed 24 metal-poor halo stars with VLT/UVES and positively detected $^6$Li
in nine of them at the $\geq 2 \sigma$ significance level.
The subdwarf LP 815-43 is the most metal-poor object (${\rm [Fe/H]} = -2.74$) with $^6$Li  in their sample.
Inoue et al. (2005) reported a tentative detection with Subaru/HDS
in the even more metal-poor star G 64-12 ([Fe/H]$=-3.17$),
although a detailed analysis is still ongoing (Aoki et al., in prep.).
With VLT/UVES, \citet{Primas00a, Primas00b} have measured the Be abundances of these two stars 
to be $\log \varepsilon ({\rm Be}) = -1.09 \pm 0.20$ for LP 815-43 and $-1.10 \pm 0.15$ for G 64-12.
The Be abundance in G 64-12 is considerably higher than that expected from
previous measurements of the Be-Fe relation in stars with similar metallicities,
and this may be the case for LP 815-43 as well.
\citet{Israelian04} analyzed nitrogen abundances in 31 metal-poor stars and found that both LP 815-43 and G 64-12
are more N rich than average.
The abundances of the relevant elements in these two stars are listed in Table \ref{tbl-obs}.
We adopt the values obtained by 1-D LTE analyses to ensure consistency.

\section{Metal-Poor SN Ic Embedded in CSM}\label{sec-SNIC}
\citet{Meynet06} simulated the evolution of metal-poor ([Fe/H]=$-6.6$, $-3.6$) massive ($60 \Msun$) stars
with rapid rotation (800 km/s) and found that C is enhanced in the outer  layers by rotation-induced mixing,
promoting intense stellar winds and significant loss of their envelopes in spite of their initial low metallicities.
At the same time, the mixing results in enrichment of N in the He layers.
If a fraction of the He layer is still remaining when the SN explodes,
it is expected that the N within is accelerated at the shock breakout,
and a significant amount of Be will be produced through the reaction ${\rm N + He} \rightarrow {\rm ^9Be}$,
thanks to its low threshold and high cross section at peak.
Thus, we consider stars that have lost all of their H layer and most of their He/N layer before the explosion.
Here we use an explosion model of a $\sim 15 \Msun$ core, originating from a main-sequence star
with mass $M_{\rm ms} \sim 40 \Msun$ \citep{Nakamura01}.
The explosion energy is assumed to be $3 \times 10^{52}$ ergs, corresponding to an energetic explosion similar to SN 1998bw.
The mass of the ejecta becomes $13 \Msun$, containing $10 \Msun$ of oxygen.
The accelerated ejecta consisting of He and CNO will collide with the circumstellar He and N stripped from the progenitor star
and produce LiBeB through the reactions ${\rm He,CNO} + {\rm He, N} \rightarrow {\rm LiBeB}$.
The energy distribution of the C/O ejecta before interaction is that calculated in \citet{Nakamura04},
shown in Figure \ref{fig-Ecs} together with the cross sections for selected reactions \citep{Read84, Mercer01}.
Rather than recalculating the explosion hydrodynamics with the added He/N layer,
we approximate by changing the composition of the accelerated outermost ejecta from C/O to He/N.
This will not lead to considerable errors as long as the replaced mass is small.

The ``thick target" approximation is used,
that is, the circumstellar He is so thick that light element production occurs entirely within the CSM
while the ejecta particles lose energy mainly by Coulomb collisions with free electrons.
This assumption is valid when the mass loss rate $\dot{M}$ is greater than $10^{-6}\,M_\odot$ yr$^{-1}$ 
for the typical wind velocity of $v_{\rm w} \sim$1,000 km s$^{-1}$, 
as can be seen by comparing the windblown material's mass column density $\sigma$ with the stopping range $R$.
We obtain for an $\alpha$-particle with initial energy $\varepsilon$
\begin{equation}
\frac{R}{\sigma} \sim 0.096 \left( \frac{\varepsilon}{10{\rm MeV/A}} \right)^2 \left(\frac{\dot{M}}{10^{-6}\,M_\odot {\rm /yr}}\right)^{-1} \left(\frac{v_{\rm w}}{1,000\,{\rm km\,s}^{-1}}\right) \left(\frac{r}{R_\odot}\right),
\end{equation}
which is well below unity. 
Here $R$ is defined as the column depth through which a particle loses all of its energy.
Even if the matter in the wind is not ionized, the energy loss rates due to ionization will be similar.
Other processes such as escape from the system can be ignored.
The LiBeB yields are calculated using the cross sections given by \citet{Read84} and \citet{Mercer01}.
We also assume that 
the composition of the innermost CSM (i.e. the last wind material) concerned with the reactions is 
the same as that of the progenitor's outermost layers.

As a result of the localized production of the light elements,
their abundance ratios with respect to heavy elements averaged over the CSM and SN ejecta
are likely to be inherited by stars of the next generation, as pointed out by \citet{Shigeyama98}.
Thus in the next section we compare the abundance ratios calculated from the above model
with the metal-poor stars discussed in the preceding section.

\section{Origin of Li and Be in Metal-poor Stars}
\begin{figure}
\begin{center}
\epsscale{.80}
\plotone{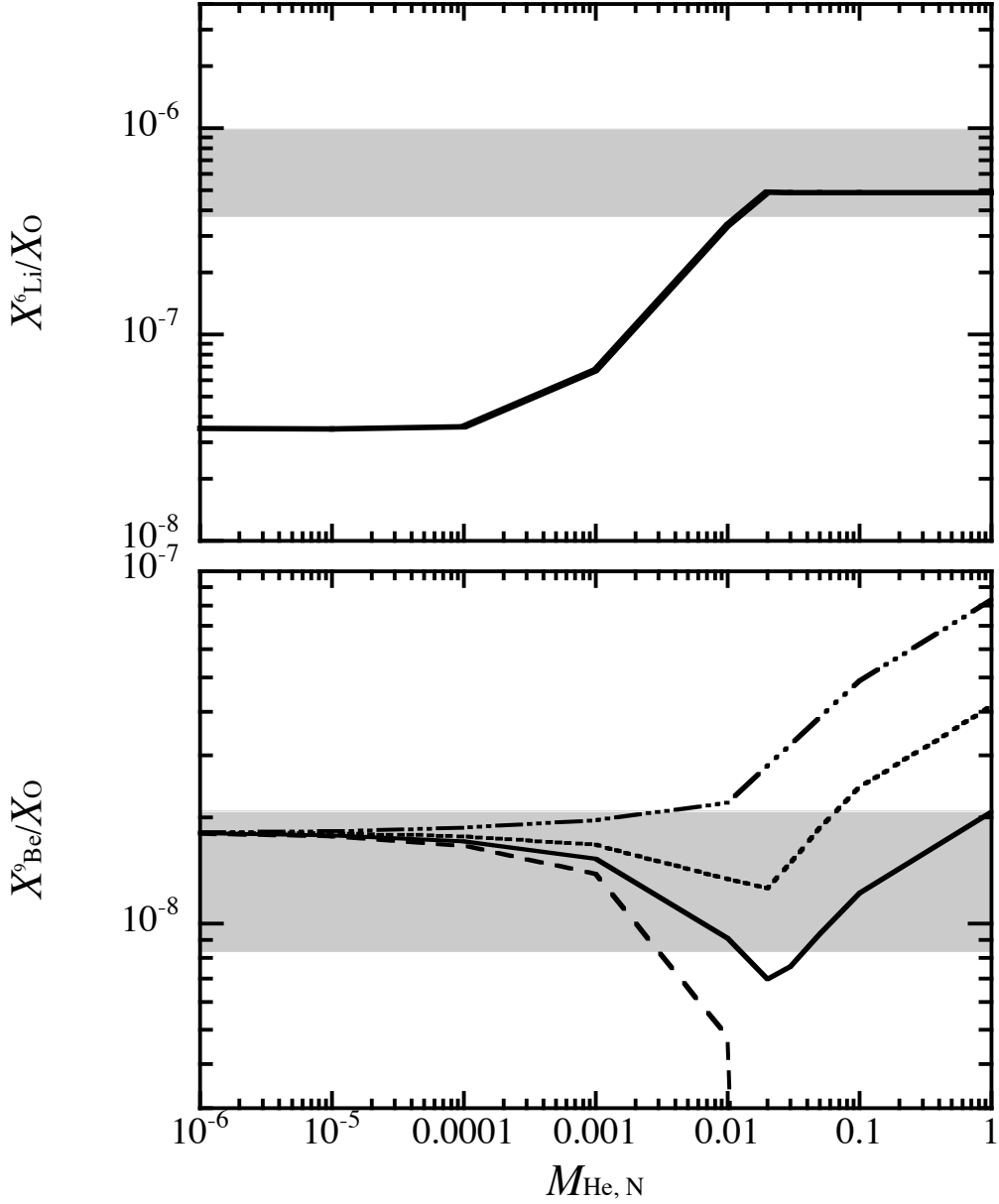}
\caption{Mass ratios of $^6$Li (top panel)  and $^9$Be (bottom) to O as functions of $M_{\rm He,\, N}$. Shown are the cases with mass fractions of N  $X_{\rm N} = 0.005$ (solid line), $X_{\rm N} = 0.01$ (dotted), $X_{\rm N} = 0.02$ (dash-dotted), and without N ($X_{\rm N}=0$; dashed). Shaded regions represent the observed ratios for the very metal-poor star LP 815-43 \citep{Primas00a, Asplund05} including the error.}
\label{fig-cal}
\end{center}
\end{figure}

We focus on the abundances of the very metal-poor star LP 815-43
from which $^6$Li, Be and N have been detected, all at apparently enhanced levels.
As seen in Table \ref{tbl-obs}, $X_{\rm ^6Li}/X_{\rm O} \sim 6.88^{+3.08}_{-3.22} \times 10^{-7}$
and $X_{\rm ^9Be}/X_{\rm O} \sim 1.32^{+0.77}_{-0.49} \times 10^{-8}$ for this star.
The mass $M_{\rm He, \, N}$ of the He/N layer on the pre-explosion core
and its mass fraction $X_{\rm N}$ of N are the main parameters.
Figure \ref{fig-cal} shows our results.
The yields of $^6$Li increase until $M_{\rm He, \, N} \sim 0.01 \Msun$, which corresponds to the threshold energy
of the ${\rm He + He} \rightarrow {\rm ^6Li} $ reaction ($\sim 11$ MeV/A, see Fig. \ref{fig-Ecs}),
and saturates for larger $M_{\rm He, \, N}$.
They depend only slightly ($\sim 0.4$\%) on $X_{\rm N}$ (as long as $X_{\rm N}$ is not very large),
because most of the $^6$Li is produced through the He + He reaction.
Only the line corresponding to the case of $X_{\rm N} = 0.005$ is shown in the top panel of Figure \ref{fig-cal}.
On the other hand, the yields of $^9$Be, which is mainly produced through the ${\rm C, O + He} \rightarrow {\rm ^9Be}$ reaction when the He/N layer is deficient, depend strongly on $X_{\rm N}$.
Without N ($X_{\rm N} = 0$), the Be yield decreases rapidly with $M_{\rm He, \, N}$,
since the ${\rm He + He}$ reaction does not produce Be.
For small $X_{\rm N}$, Be yields first decrease with $M_{\rm He, \, N}$ for the above reason,
and then turn to increase since the cross section of the ${\rm N + He} \rightarrow {\rm ^9Be}$ reaction
peaks around $\sim 13$ MeV/A, corresponding to $\sim 0.013\Msun$ integrated from outside
for SN 1998bw model (see Fig. \ref{fig-Ecs}). 
It should be noted that if the He layer is sufficiently large, for instance $M_{\rm He} \gtrsim 1 \Msun$,
our results here tend to overestimate the light element yields because a core with such a large He layer
might not be so compact in reality and effective acceleration at the shock breakout is not expected.
To specify how much He can be accelerated above the production threshold in such cases,
detailed calculations for the structures of mass-losing stars are required.

Both ${\rm X_{^6Li}/X_O}$ and ${\rm X_{^9Be}/X_O}$ show good agreement with the observational data of LP 815-43
when $M_{\rm He, \, N} \sim 0.01-0.1 \Msun$ and $X_{\rm N} \sim 0.005-0.01$.
These are consistent with recent simulations of the evolution of metal-poor massive stars with rotation:
$X_{\rm N} \sim 0.008$ for $M_{\rm ms} = 40 \Msun$ (Hirschi et al., in preparation)
and $X_{\rm N} \sim 0.01$ for $M_{\rm ms} = 60 \Msun$ \citep{Meynet06}.
The observed ${\rm X_{^9Be}/X_O}$ for G64-12 can also be reproduced
with parameters analogous to those for LP 815-43.
The isotopic ratio ${\rm ^6Li/^7Li}$ in our models lies between 0.27 and 1.4,
which is much larger than the observed value $0.046 \pm 0.022$ for LP 815-43 \citep{Asplund05}
so that the scenario discussed here is not expected to contribute significantly to $^7$Li.

\section{Conclusions and Discussion}\label{sec-disc}
In this Letter we have proposed a new mechanism to produce the light elements,
especially $^6$Li and Be recently observed in metal-poor stars.
This is based on recent theoretical findings that rotating, metal-poor massive stars
can lose substantial amounts of their envelopes,
and end up with SNe Ic with a small amount of He and N in the outermost layers.
The outer layers composed of He and CNO are accelerated at the shock breakout
and undergo spallation and He-He fusion reactions to produce mainly $^6$Li and $^9$Be.
Our calculations show that if $\sim 0.01-0.1\,M_\odot$ of the He layer is remaining on the core
in a SN explosion of a massive star with $M_{\rm ms}>30\,M_\odot$,
then the observed abundance ratios $^6$Li/O can be reproduced.
If the He layer contains a small amount ($\sim 0.5 - 1$\%) of N, we can also reproduce $^9$Be/O. 

\begin{figure}
\begin{center}
\epsscale{.80}
\plotone{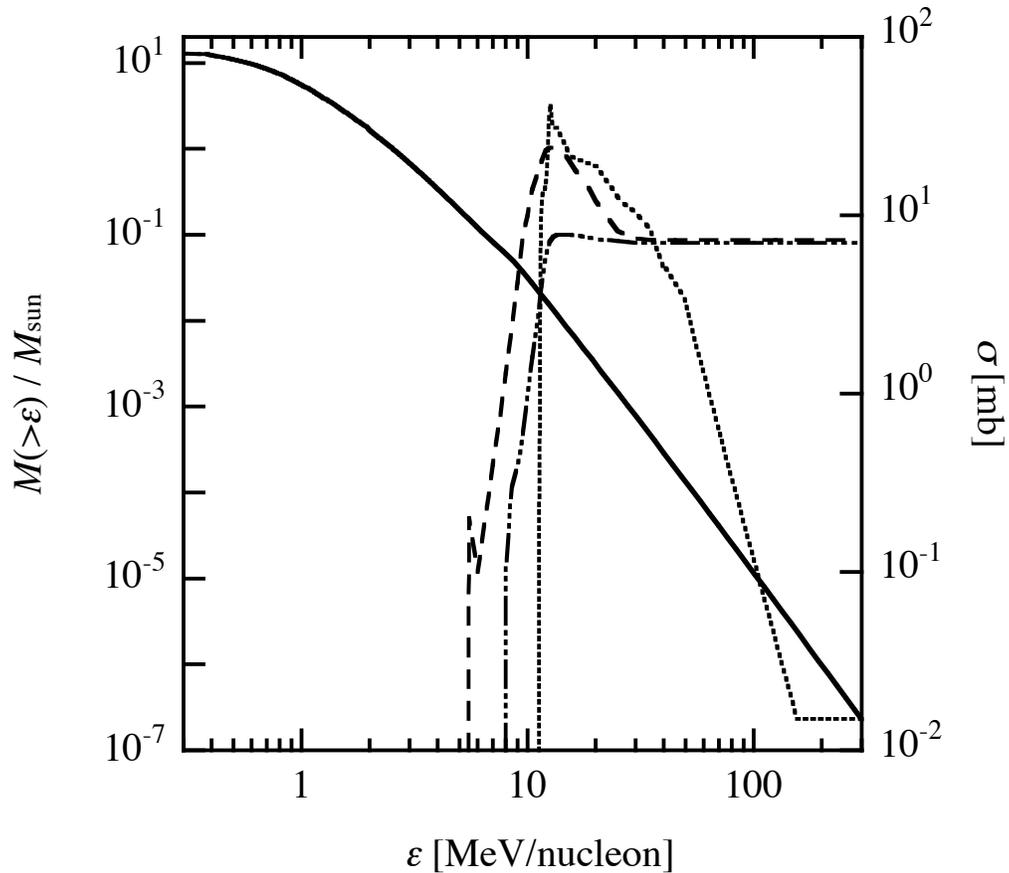}
\caption{The energy distribution of ejecta and cross sections of reactions as functions of energy per nucleon $\varepsilon$. The solid line represents the result of a numerical calculation \citep{Nakamura04} where $M(>\varepsilon)$ denotes the mass of ejecta that have particle energy per nucleon greater than $\varepsilon$. The dotted, dashed, and dash-dotted lines show the cross sections $\sigma$ of reactions ${\rm He + He} \rightarrow {\rm ^6Li}$,  ${\rm N + He} \rightarrow {\rm ^9Be}$, and ${\rm O + He} \rightarrow {\rm ^9Be}$, respectively.}
\label{fig-Ecs}
\end{center}
\end{figure}

Compared with the standard picture of cosmic ray shock acceleration by normal supernovae,
our high yield of $^6$Li results from three crucial differences.
First, we consider a large explosion energy, of order $10^{52}$ erg, as observed in some SNe such as SN 1998bw.
Second, our mechanism considers the direct interaction of fast ejecta with the CSM,
and does not involve an efficiency factor for cosmic ray acceleration
and is not affected by losses or escape during ISM propagation.
Third, the energy distribution of the accelerated particles is a very steeply falling power-law,
with spectral index $\sim 4.6$ as opposed to $2$ for shock accelerated particles.
This means that most of the energy is contained in the lowest energy portions,
where the cross sections for both ${\rm He + He} \rightarrow {\rm ^6Li}$ and  ${\rm N + He} \rightarrow {\rm ^9Be}$ peak
(see Fig. \ref{fig-Ecs}).
In fact, the total energy of particles above 10 MeV/nucleon in our fiducial model
is $8\times10^{50}$ erg, much higher than is achievable in the standard picture, 10-30\% of $10^{51}$ erg.

One may claim that $^6$Li, and probably also $^9$Be \citep{GPerez05},
in metal-poor stars may have been depleted from their initial values,
since the observed abundances of $^7$Li/H are a few times smaller than that predicted by standard big-bang nucleosynthesis,
and $^6$Li is more fragile than $^7$Li \citep{Asplund05}.
The actual survival fraction of $^6$Li is difficult to evaluate reliably 
because the pre-main-sequence Li destruction is sensitive to convection,
which cannot be modeled without free parameters.
However, depletion factors as  large as 10 may still be compatible with our picture.
The estimates in the preceding section is based on an explosion energy of the order of $10^{52}$ ergs.
The mass of ejecta with energy per nucleon above a certain value scales very strongly with the explosion energy,
$E_{\rm ex}$ as $E_{\rm ex}^{3.4}$ \citep{Nakamura04}.
Thus a SN with the explosion energy 2 times higher than that considered in the preceding section
will be able to produce $\sim 10$ times larger amounts of light elements depending on the distribution of He and N.
Nevertheless, if such very energetic explosions turn out to be rare,
the mechanism proposed here may play a rather limited role
compared to other potential LiBeB production processes.

Both the degree of mass loss and the amount of N enrichment in the progenitor star
are expected to be sensitive to its initial rotation speed \citep{Meynet02, Meynet06}.
The actual rotation speed is presumably distributed over a wide range,
as is the explosion energy and the extent of mass loss at the time of the explosion.
Therefore, dispersions in the $^6$Li and Be abundances are expected,
and current observations suggest that this may indeed be the case
for $^6$Li \citep{Aoki04,Asplund05,Inoue05} as well as Be \citep{Boesgaard05}.
Our scenario predicts a close relation between Be and N, since Be arises directly as a consequence of N spallation.
A looser correlation between $^6$Li and N is also expected,
as effective $^6$Li production requires sufficient mass loss, which in turn implies significant N enrichment.
Further observations of $^6$Li, Be and N for a large sample of metal-poor stars should provide definitive tests.
Note that such correlations are not expected for other scenarios which involve mainly $^6$Li production only,
such as structure formation cosmic rays \citep{Suzuki02,Rollinde05},
active galactic nuclei outflows \citep{Prantzos06,Nath06},
and production processes in the early universe \citep{Jedamzik05}.
It is also mentioned that mass loss onto companion stars in binary systems may represent
an additional pathway to our scenario, provided that
sufficiently thick CSM is remaining at the time of the explosion.
We reiterate that more observational data is necessary to elucidate
what fraction of the LiBeB abundances seen in halo stars of different metallicity
can be explained by the mechanism proposed here.

We are grateful to Georges Meynet, Raphael Hirschi, Takeru K. Suzuki and Sean Ryan for valuable discussions.
We also acknowledge the contributions of an anonymous referee,
which led to clarification of a number of points in our original manuscript.
This work has been partially supported by the grant in aid (16540213, 17740108) of the Ministry of Education, Science, Culture, and Sports in Japan.

\clearpage
\begin{deluxetable}{lcccccc}
\tabletypesize{\scriptsize}
\tablecaption{Abundances from the literature. \label{tbl-obs}}
\tablewidth{0pt}
\tablehead{
 \colhead{Star} & \colhead{[Fe/H]} & 
\colhead{$\log \varepsilon ({\rm ^7Li})$}   & \colhead{${\rm ^6Li/^7Li}$}   & 
\colhead{$\log \varepsilon ({\rm Be})$} &\colhead{$\log \varepsilon ({\rm N})$}   &
\colhead{$\log \varepsilon ({\rm O})$}
}
\startdata
LP 815-43   & $-$2.74\tablenotemark{a} & 2.16\tablenotemark{b} & 
0.046$\pm$0.022\tablenotemark{b} & $-$1.09$\pm$0.20\tablenotemark{c} & 
5.61\tablenotemark{a} & 6.54\tablenotemark{c}\\
G 64-12      & $-$3.17\tablenotemark{d} & 2.30\tablenotemark{e} & 
- & $-$1.10$\pm$0.15\tablenotemark{e} &     
6.15\tablenotemark{a} & 6.34\tablenotemark{d}\\

 \enddata

\tablenotetext{a}{\citet{Israelian04}}
\tablenotetext{b}{\citet{Asplund05}}
\tablenotetext{c}{\citet{Primas00a}}
\tablenotetext{d}{\citet{Akerman04}}
\tablenotetext{e}{\citet{Primas00b}}

\end{deluxetable}

\clearpage

\clearpage
\clearpage

\end{document}